\documentclass[journal]{IEEEtran}
\usepackage[utf8]{inputenc}
\usepackage[colorlinks=true,linkcolor=blue,citecolor = blue]{hyperref}%

\usepackage{hyperref}
\hypersetup{
    colorlinks=true,
    linkcolor=blue,
    filecolor=magenta,      
    urlcolor=blue,
}

\usepackage[sort, numbers]{natbib}
\usepackage{graphicx}
\usepackage{xr}
\usepackage[dvipsnames]{xcolor}
\colorlet{Mycolor1}{green!10!orange!90!}
\colorlet{Mycolor2}{green!30!black!70!}
\colorlet{Mycolor3}{blue!70!red!30!}
\usepackage{textcomp}
\usepackage{amsmath}
\usepackage{amssymb}
\usepackage{enumitem}
\usepackage[flushleft]{threeparttable}
\renewcommand{\vec}[1]{\mathbf{#1}}
\usepackage{tikz}

\usepackage{tikz}
\usetikzlibrary{tikzmark}

\newcommand{\highlight}[1]{    {\textcolor{black}{#1}}   }
\newcommand{\highli}[1]{    {\textcolor{black}{#1}}   }
\newcommand{\high}[1]{    {\textcolor{black}{#1}}   }
\makeatletter
\newcommand{\thickhline}{%
    \noalign {\ifnum 0=`}\fi \hrule height 1pt
    \futurelet \reserved@a \@xhline
}
\ifCLASSINFOpdf
\else
\fi
\newcommand\copyrighttext{%
  \footnotesize \textcopyright 2020 IEEE.  Personal use of this material is permitted.  Permission from IEEE must be obtained for all other uses, in any current or future media, including reprinting/republishing this material for advertising or promotional purposes, creating new collective works, for resale or redistribution to servers or lists, or reuse of any copyrighted component of this work in other works.
  DOI: \href{https://ieeexplore.ieee.org/document/9222178}{10.1109/JIOT.2020.3030492}}
  
\newcommand\copyrightnotice{%
\begin{tikzpicture}[remember picture,overlay]
\node[anchor=south,yshift=1pt] at (current page.south) {\fbox{\parbox{\dimexpr\textwidth-\fboxsep-\fboxrule\relax}{\copyrighttext}}};
\end{tikzpicture}%
}

\newcommand\copyrighttexttop{%
  \footnotesize This paper has been accepted for publication in the IEEE Internet of Things Journal on 08-Oct-2020,
  DOI: \href{https://ieeexplore.ieee.org/document/9222178}{10.1109/JIOT.2020.3030492}}
  
\newcommand\copyrightnoticetop{%
\begin{tikzpicture}[remember picture,overlay]
\node[anchor=north,yshift=1pt] at (current page.north) {\fbox{\parbox{\dimexpr\textwidth-\fboxsep-\fboxrule\relax}{\copyrighttexttop}}};
\end{tikzpicture}%
}

\begin{document}
%
\title{Is Image Encoding Beneficial for Deep Learning in Finance?\\
\large An Analysis of Image Encoding Methods for the Application of Convolutional Neural Networks in Finance}
%
%
%

\author{
        Dan~Wang,
        Tianrui~Wang,
        and Ionu\c{t}~Florescu,
\thanks{Manuscript submitted July 8, 2020. This work is supported by UBS Group}
\thanks{Dan~Wang is with the School of Business, Stevens Institute of Technology, Hoboken,
NJ, 07030 USA (e-mail: dwang35@stevens.edu.)}
\thanks{Tianrui~Wang is with the School of Business, Stevens Institute of Technology, Hoboken,
NJ, 07030 USA (e-mail: twang55@stevens.edu.)}
\thanks{Ionu\c{t}~Florescu is with the School of Business, Stevens Institute of Technology, Hoboken,
NJ, 07030 USA (e-mail: ifloresc@stevens.edu.)}
}

\maketitle
\copyrightnotice
\copyrightnoticetop
\begin{abstract}
\highlight{In 2012, SEC mandated all corporate filings for any company doing business in US be entered into the Electronic Data Gathering, Analysis, and Retrieval (EDGAR) system. In this work we are investigating ways to analyze the data available through EDGAR database. This may serve portfolio managers (pension funds, mutual funds, insurance, hedge funds) to get automated insights into companies they invest in, to better manage their portfolios. The analysis is based on Artificial Neural Networks applied to the data.} In particular, one of the most popular machine learning methods, the Convolutional Neural Network (CNN) architecture, originally developed to interpret and classify images, is now being used to interpret financial data. This work investigates\highlight{the best way to input data collected from the SEC filings into a CNN architecture.}
We incorporate accounting principles and mathematical methods into the design of three image encoding methods. Specifically, two methods are derived from accounting principles (Sequential Arrangement, Category Chunk Arrangement) and one is using a purely mathematical technique (Hilbert Vector Arrangement). In this work we analyze fundamental financial data as well as financial ratio data and study companies from the financial, healthcare and IT sectors in the United States. We find that using imaging techniques to input data for CNN works better for financial ratio data but is not significantly better than simply using the 1D input directly for fundamental data. We do not find the Hilbert Vector Arrangement technique to be significantly better than other imaging techniques.   
\end{abstract}


%
\IEEEpeerreviewmaketitle

\section{Introduction}
%
%
%
%
\IEEEPARstart{I}{n} finance, Generally Accepted Accounting Principles (GAAP) is a common set of accounting principles, standards, and procedures that companies in the U.S. must follow when they compile and submit their financial statements. 
Researchers and rating agencies assess the credit worthiness of corporations based on these financial statements,\highlight{and investors rebalance their portfolios based on updated credit worthiness.}
\highlight{Thus, understanding} the fundamental relationship between accounting variables, financial ratios and corporate credit rating 
\highlight{can help investors take informed decisions on modifying their portfolio.}

Corporate financial reports are hard to analyze. Most rating agencies use a combination of relatively simple models with a lot of expertise assessments, as it is inherently difficult to associate hundreds of features contained in financial statements with a single credit rating score. For instance, the Compustat\textregistered{} dataset \citep{SP2019compustat}, used to research US public companies, contains $332$ financial accounting variables collected from the original financial statements. Immediate questions come to our mind: among the $332$ are there any features significantly more important than others? Are those ``important'' features common for all industry sectors, or does the best feature set vary by sectors? 

Feature selection could be one of the techniques used to answer these questions. The goal of feature selection is to reduce the dimensionality of the input variable space and to improve the efficiency of statistical or learning models \citep{deegalla2006reducing}. Therefore, it is used to eliminate redundant, irrelevant, and noisy parts of the input data. It is generally argued that this process can improve the performance of classification models efficiently \citep{yang2006dimensionality, cannas2013framework}. In financial literature, the feature selection problem has been addressed using several techniques. For example, Principal component analysis (PCA) \citep{joliffe1992principal}, Chi Square testing \citep{jin2006machine}, genetic algorithm \cite{leardi2000application, frohlich2003feature}, and the Gini index \citep{deng2012feature}.

In our previous projects replicating these methods on current and updated datasets \citep{Apeksha2018chi, golbayani2020application, qi2018genetic}, we obtain some interesting results on the feature selection problem. We find that some features are indeed more important than others, but the set of the so called ``best features'' varies from sector to sector. We note from our tests that the same feature selection technique provides a different set of best features. We exemplify by comparing the best set of features for a Support Vector Machine (SVM) classification versus the best features for a Multilayer Perceptron (MLP) algorithm. 

In particular, we analyze the feature selection problem for credit rating using only neural networks as classifiers in \cite{golbayani2020application}. One of the conclusions is that, for the credit rating problem, it is better to use a convolutional layer in the CNN architecture on all available variables rather than using a traditional MLP on the best feature subset. We are not the first to notice the advantages of the CNN architecture. Indeed, CNN is shown to improve the performance of classification problems due to its feature learning capability \citep{li2017deep,jing2017convolutional}. 

This prior work motivates the current article. Since the best approach seems to use all the available features in a CNN architecture, does it make a difference how the features are inputted?

\highlight{A typical CNN architecture will take advantage of the spatial relationship between neighboring pixels corresponding to parts of an object in an image. In a set of financial statements, a company's fundamental data \citep{SP2019compustat} is categorized in balance sheets, income statements, and statements of cash flow. The features in the same category have closer relationship than features between different categories. For example, features in the balance sheet show what a company owns (assets) and owes (liabilities) at a specific moment, while features in the income statement show total revenues and expenses for a period of time. Lenders use the balance sheet to see if they should extend any more credit, and they use the income statement to decide on whether or not the business is making enough profit to pay its liabilities.}

\highlight{Furthermore, a lot of accounting variables are closely related. For example, it is commonly known that assets equal to the sum of liabilities and equity; and assets could be broken down into current assets and non-current assets. This is very similar to pixels forming up a particular object in an image. Table \ref{tab:balancesheet} shows a small section of a sample balance sheet using data from the Compustat\textregistered{} dataset, which categorizes the accounting variables into primary, secondary and minor categories, and calculates totals and sub-totals. }

\highlight{In this work we examine whether using this hierarchy of accounting variables may improve the performance of the rating models. Could we arrange the features according to our knowledge of fundamental finance and accounting to produce better results?}

\begin{table}[htbp]
  \centering
  \caption{A Sample Section of A Balance Sheet}
    \begin{tabular}{rlrr}
    \hline
    \multicolumn{4}{c}{BALANCE SHEET XXX LLC} \\
    \multicolumn{4}{c}{AS OF FY17 Q2} \\
    \multicolumn{4}{c}{(in millions)} \\
    \hline
    \multicolumn{1}{l}{\textbf{Current Assets - Total}} &       & \textbf{\$8,468 } &  \\
    \multicolumn{1}{l}{\quad Cash and Short-Term Investments} &       & \underline{4,454} &  \\
    \multicolumn{1}{l}{\quad \quad Cash }  &   &       & 4,254 \\
    \multicolumn{1}{l}{\quad \quad Short-Term Investments- Total }      &  &       & 200 \\
    \multicolumn{1}{l}{\quad Receivables - Total} &       & \underline{2,073} &  \\
    \multicolumn{1}{l}{\quad \quad Receivables (Net) - Utility} & &       & 0 \\
    \multicolumn{1}{l}{\quad \quad Receivables - Trade}  &  &       & 2,073 \\
    \multicolumn{1}{l}{\quad \quad Receivables - Current Other incl Tax Refunds}  &  &       & 0 \\
    \multicolumn{1}{l}{\quad \quad Unbilled Receivables - Quarterly}  &  &       & 0 \\
    \multicolumn{1}{l}{\quad \quad Receivables - Estimated Doubtful}  &  &       & 0 \\
    \multicolumn{1}{l}{\quad Inventories - Total} &       & \underline{1,311} &  \\
    \multicolumn{1}{l}{\quad \quad Inventories}  &  &       & 0 \\
    \multicolumn{1}{l}{\quad \quad Inventory - Raw Materials}  &  &       & 197 \\
    \multicolumn{1}{l}{\quad \quad Inventory - Work in Process}  &  &       & 632 \\
    \multicolumn{1}{l}{\quad \quad Inventory - Finished Goods}  &  &       & 482 \\
    \multicolumn{1}{l}{\quad \quad Inventory - Other}  &  &       & 0 \\
    \multicolumn{1}{l}{\quad Current Assets - Other - Total} &       & \underline{630}   &  \\
    \multicolumn{1}{l}{\quad \quad Current Assets - Other - Utility}  &  &       & 0 \\
    \multicolumn{1}{l}{\quad \quad Current Deferred Tax Asset}  &  &       & 0 \\
    \hline
    \end{tabular}%
  \label{tab:balancesheet}%
\end{table}%

Unfortunately, financial datasets, particularly fundamental datasets are one dimensional data. The dichotomy of applying techniques developed for 2 dimensional spatial objects to 1 dimensional vectors is the main object of our investigation.  \highlight{The work detailed herein may be used as a data monetization process through the free EDGAR database\footnote{\highlight{The U.S. Securities and Exchange Commission (SEC) describes EDGAR as follows: ``EDGAR, the Electronic Data Gathering, Analysis, and Retrieval system,
performs automated collection, validation, indexing, acceptance, and forwarding of submissions by companies and
others who are required by law to file forms with the U.S. Securities and Exchange Commission (SEC). Its primary
purpose is to increase the efficiency and fairness of the securities market for the benefit of investors, corporations, and
the economy by accelerating the receipt, acceptance, dissemination, and analysis of time-sensitive corporate information
filed with the agency.''}}.
The analytics developed may help improve investment decisions made by asset holders, such as mutual and pension funds, and generally portfolio managers.} 

In literature there are several ways to implement the CNN architecture to financial data. The first approach is to use the vector of inputs directly into the convolutional layer. Reference \cite{kvamme2018predicting} uses the raw account transaction data of a borrower directly to predict mortgage default. This architecture uses a $1 \times N$ vector as input. We will refer this architecture as 1D CNN. The author of \cite{zhao2017convolutional} proposes the same 1D CNN classification model for a time-series data structure. 

The second approach used in literature is to encode the data into a two dimensional vector (matrix). This technique is used for variables collected in time (e.g., minute sampled asset price data). In \cite{hatami2018classification} the authors use the so called Recurrence Plots method to encode this type of time series data into a 2 dimensional image. The image is then used as input into a CNN model used for classification. In \cite{wang2015encoding} the authors discuss and compare two methods (Gramian Angular field (GAF) and Markov Transition Field (MTF)) to encode time-series data into an image. The authors find that MTF has higher error rate for classification under CNN framework. References \cite{gudelek2017deep} and \cite{sezer2018algorithmic} construct an image from technical analysis indicators, and propose a trading strategy using a CNN model to predict stock price movement.

These methods are generally aggregating daily or high frequency data for a certain period to create an image that may serve as input into the CNN. We try to apply this idea to credit rating in \cite{golbayani2020application}. In that paper, we create a $4\times 332$ input image from the last $4$ quarters each consisting of $332$ financial variables. Although this architecture produces better results than using a single quarter in a 1D CNN architecture, we find that the performance is inferior to a specifically designed recurrent architecture such as Long Short-term Memory (LSTM). 

\highlight{We note that the techniques mentioned in the previous citations are all applied on temporal data sampled with high frequency. This is difficult to implement in the credit rating assessment. In our previous work we used the past four 10-Q quarterly statements to assess a company's rating. 10-Q statements are mandatory for all publicly traded companies. However, quarterly statements are not widely required by countries other than the U.S. Private companies in the U.S. and companies around the world issue financial statements annually. Thus, for credit rating we generally do not have enough temporal data to create an image from several quarters. Therefore, we investigate the feasibility of using just the most recent statement to assess corporate credit rating.}

In this work, we are investigating whether grouping variables from a financial statement in a manner similar to pixels in an image may maximize the performance of Convolutional Neural Networks to assess credit rating. Our primary guide to determine the nature of financial variables is the FASAB Handbook of Federal Accounting Standards and Other Pronouncements issued by \cite{federal2014fasab}.  

Moreover, several recent papers \cite{kavitha2017deep, yin2018image, loka2019hilbert} apply a similar technique to 1 dimensional DNA feature vector and transform it into a two dimensional image format. Specifically, the authors adopt a Hilbert space filling-curve for feature encoding \cite{moon2001analysis} and combine it with a CNN architecture to assess its performance. The authors argue that this type of encoding is superior to any other way to transform 1D data into a 2D image in terms of maintaining close relationship between neighbor features and enlarging distance between unrelated features. We implement this encoding technique to financial data, and then analyze and compare its performance with other types of encoding data into an image. 

Section \ref{sec:methodology} presents the encoding methods used and provides a brief review of the CNN architecture. We use the variables presented in the Compustat\textregistered{} database \citep{SP2019compustat}. Section \ref{sec:results} presents experimental results obtained by applying the encoding methodology on different datasets.

\section{Methodology\label{sec:methodology}}

In this section we describe the encoding methods and the convolutional neural network architecture. 


\subsection{Data}

In this study, we use two datasets: the quarterly fundamental data and the financial ratio data both obtained from the Compustat\textregistered{}  Database \citep{SP2019compustat}. The fundamental dataset contains 332 accounting variables while the financial ratio dataset contains 69 variables. In this work we focus on financial, healthcare and information technology (IT) sectors in the US market. We study the time interval $2000-2016$, and we have $66$ , $59$ and $69$ companies in each sector chosen, respectively. 

\begin{table}[htbp]
  \centering
  \tabcolsep=1.3pt\relax
    \footnotesize
  \caption{Rating mapping}
    \begin{tabular}{c|cccccc}
    \hline
    Original Rating & AAA   & \multicolumn{1}{c|}{AA+} & AA    & \multicolumn{1}{c|}{AA-} & A+    & A \\
    New Scheme & 0     & \multicolumn{1}{c|}{0} & 1     & \multicolumn{1}{c|}{1} & 2     & 3 \\
    Rating description & \multicolumn{2}{c|}{Prime} & \multicolumn{2}{c|}{High grade} & \multicolumn{2}{c}{Upper medium } \\
    \hline
    Original Rating & \multicolumn{1}{c|}{A-} & BBB+  & BBB   & \multicolumn{1}{c|}{BBB-} & BB+   & BB \\
    New Scheme & \multicolumn{1}{c|}{4} & 5     & 6     & \multicolumn{1}{c|}{7} & 8     & 9 \\
    Rating description & \multicolumn{1}{c|}{grade} & \multicolumn{3}{c|}{Lower medium grade} & \multicolumn{2}{c}{Non-investment grade } \\
    \hline
    Original Rating & \multicolumn{1}{c|}{BB-} & B+    & B     & \multicolumn{1}{c|}{B-} & CCC+  & CCC \\
    New Scheme & \multicolumn{1}{c|}{9} & 10    & 10    & \multicolumn{1}{c|}{10} & 11    & 11 \\
    Rating description & \multicolumn{1}{c|}{speculative} & \multicolumn{3}{c|}{Highly speculative} & \multicolumn{2}{c}{Extremely risks} \\
    \hline
    Orignial Rating & CCC-  & CC    & C     & D     & SD    & N.M. \\
    New Scheme & 11    & 11    & 11    & 11    & 11    & 11 \\
    Rating description & \multicolumn{6}{c}{Extremely risks} \\
    \hline
    \end{tabular}%
  \label{tab:ratingmap}%
\end{table}%

\begin{figure}[htp]
    \centering
    \includegraphics[width=8.8cm]{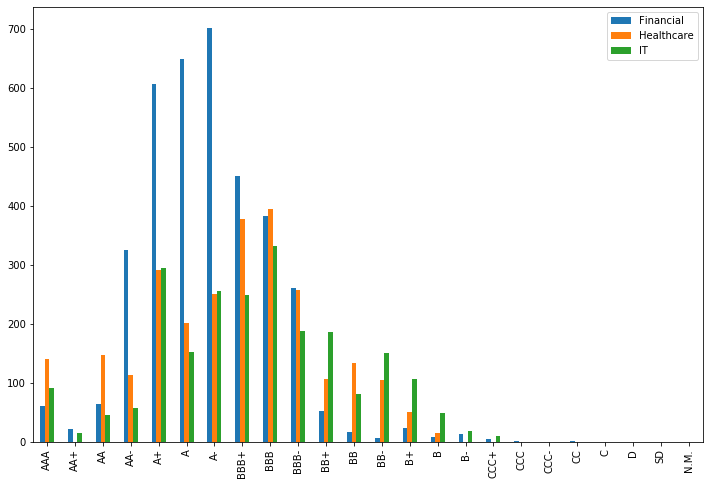}
    \caption{Original Rating distribution }
    \label{fig:map_before}
\end{figure}

\begin{figure}[htp]
    \centering
    \includegraphics[width=8.8cm]{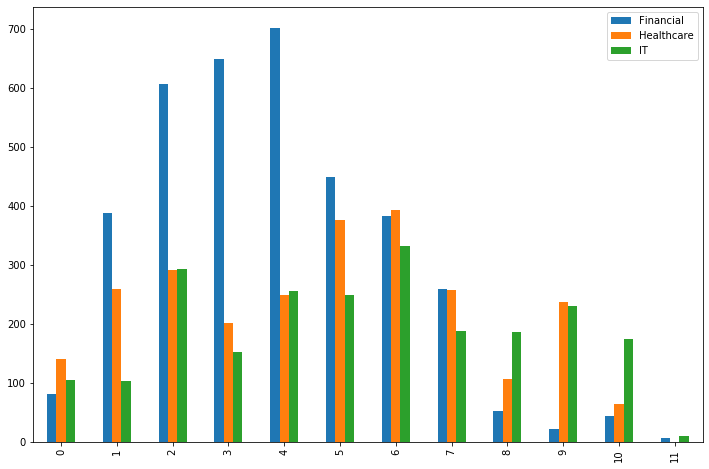}
    \caption{rating distribution in new scheme}
    \label{fig:map_after}
\end{figure}

We use Standard and Poor's credit ratings as the benchmark. The distribution of the original rating scheme is extremely unbalanced. This unbalanced distribution generally hinders machine learning performance. To balance out the distribution, we design a new scheme, relevant from an investing perspective.  We cite \cite{langohr2010rating} which creates a categorization on investment instruments (line 3 in Table \ref{tab:ratingmap}). Starting with the two schemes and analyzing the available data we use a mapping scheme based on the risk category of rating. This scheme is presented in line 2 of Table \ref{tab:ratingmap}. To make the rating distribution of new scheme relatively balanced, we keep all ratings in upper medium, lower medium, and non-investment grade. We could see the original distribution in Figure \ref{fig:map_before}, as well as the distribution based on the new rating in Figure \ref{fig:map_after}.

In our previous study \cite{golbayani2020application} we show that using a random allocation of data points among training and test data could artificially increase the results obtained by as much as 10\%. In that paper we conclude an out-of-time allocation between training and test data is more realistic for financial time series data. Thus, in this work we adopt an out-of-time test method rather than an out-of-sample test. Specifically, the test set is obtained by holding out one year from the remaining data which constitutes the training set. 
\highlight{Consequently, the training data set contains 332 features for the fundamental data and 69 features in the financial ratio data ranging from 2000 to 2015, while the testing data set contains 4 quarters from 2016.}

\subsection{Input data encoding methods}
\subsubsection{Data Re-structuring\label{sec:dataRestructuring}}
The Compustat\textregistered{}  fundamental data \citep{SP2019compustat} includes 332 features extracted from financial statements of public companies, ordered alphabetically by the name of the features. In this work we want to take into consideration the sequential relationships and hierarchy among accounting features, as one feature could be a simple combination of the previous two or more features. Therefore, we re-structure all the features to comply with the format and rules of a typical set of financial statements defined in \cite{federal2014fasab}:
\begin{enumerate}[label=\alph*)]
    \item A Beginning of the Period Statement of Financial Position;
    \item An End of the Period Statement of Financial Position; 
    \item A Statement of Earnings and Comprehensive Income;
    \item{\label{statementchanges}}  A Statement of Changes in Equity;
    \item\label{statementcash}  A Statement of Cash Flows.
\end{enumerate}
Here, a and b are known as the  balance sheet, and c is known as the income statement.
We do not include the Statement of Changes in Equity \ref{statementchanges} or the Statement of Cash Flows \ref{statementcash} into our categorization process. Most of the accounting items included in these two statements can also be found in the income statement or the balance sheet, and thus be duplicated in the feature space.

In the process of compiling the financial report, many supplemental items are used to calculate the accounting items in the balance sheet and the income statement. These items are also included in the dataset given by \cite{SP2019compustat}, and are classified as ``Balance Sheet Supplemental Data'' and ``Income Statement Supplemental Data'' respectively, according to the Online Manual. Moreover, there are other items that are not covered by either the balance sheet or the income statement. Referring to the online manual issued by \cite{SP2019compustat}, we introduce two more sections called ``Special Items'' and ``Standard \& Poor's Core Earnings™'' to cover these items.

Therefore, the set of $332$ features is divided into 6 final sections in the order shown in Table.~\ref{tab_AccountingFeature}. The sections are compiled using the FASAB Handbook of Federal Accounting Standards and Other Pronouncements issued by \cite{federal2014fasab} and the Online Manual by \cite{SP2019compustat}.

\begin{table}[h!]
  \centering
  \caption{Six Sections of Accounting Features}
    \begin{tabular}{lr}
    Section & \multicolumn{1}{l}{Feature Count} \\
    \hline
    Balance Sheet Data & 78 \\
    Balance Sheet Supplemental Data & 45 \\
    Income Statement Data & 75 \\
    Income Statement Supplemental Data & 33 \\
    Special Items & 49 \\
    {Standard \& Poor's Core Earnings™} & 50 \\
    \hline
    \\
    \end{tabular}%

  \label{tab_AccountingFeature}%
\end{table}%


\highlight{Besides the fundamental data described above, we use a financial ratio data set available from the Compustat\textregistered{}  database. These ratios are derived from the original fundamental data and are used by financial professionals to better understand a company's financial state. We use $69$ such ratios classified by \cite{SP2019compustat} into $8$ categories. Table \ref{RatioFeature} shows the feature counts for the eight categories. According to \cite{wrds2016WIFR}, these categories are:
\begin{enumerate}
    \item \textit{Valuation}: estimates the attractiveness of a firm’s stock (overpriced or underpriced), e.g.: P/E ratio, Shiller’s CAPE ratio;
    \item \textit{Profitability}: measures the ability of a firm to generate profit, e.g.: ROA, Gross Profit Margin;
    \item  \textit{Capitalization}: measures the debt component of a firm’s total capital structure, e.g.: Capitalization Ratio, Total Debt-to-Invested Capital Ratio;
    \item \textit{Financial Soundness}: captures the firm’s financial healthiness, e.g.: Receivables to Current Assets, Cash Flow to Total Debt;
    \item \textit{Solvency}: captures the firm’s ability to meet long-term obligations, e.g.: Total Debt to Equity Ratio, Interest Coverage Ratio;
    \item \textit{Liquidity}: measures a firm’s ability to meet its short-term obligations, e.g.: Current Ratio, Quick Ratio;
    \item  \textit{Efficiency}: captures the effectiveness of firm’s usage of assets and liability, e.g.: Asset Turnover, Inventory Turnover;
    \item \textit{Others}: Miscellaneous ratios, e.g.: R\&D-to-Sales, Labor Expenses-to-Sales.
\end{enumerate}
}
\begin{table}[htbp]
  \centering
  \caption{Eight Categories of Ratio Features}
    \begin{tabular}{lr}
    Category & \multicolumn{1}{l}{Feature Count} \\
    \hline
    Validation & 13 \\
    Profitability & 15 \\
    Capitalization & 4 \\
    Financial Soundness & 16 \\
    Solvency & 6 \\
    Liquidity & 4 \\
    Efficiency & 7 \\
    Other & 4 \\
    \hline
    \\
    \end{tabular}

  \label{RatioFeature}
\end{table}

\subsubsection{Data Encoding}
A typical approach for changing data dimensionality of a vector would be to transform a $n\times 1$ vector into a two dimensional matrix. One of the underlying principles we wanted to follow is to group similar features together in an ``objects in an image'' idea. 

\begin{figure}[htbp]
\centerline{\includegraphics[width=8.8cm]{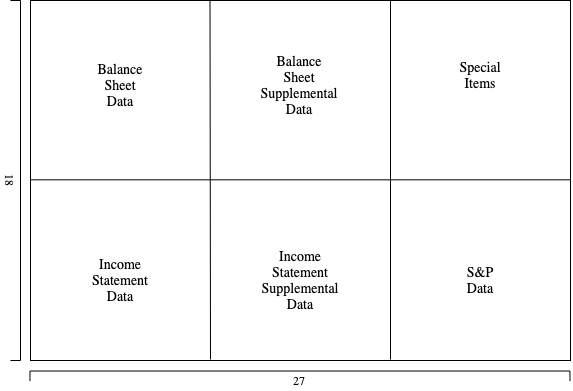}}
\caption{Categorical Chunk Arrangement For Financial Data\label{Categorical Chunk Arrangement}}
\end{figure}

\begin{figure}[htbp]
\centerline{\includegraphics[width=8.8cm]{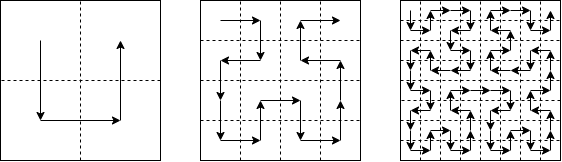}}
\caption{Hilbert Curve\label{Hilbert Curve}}
\end{figure}

We first list all the types of ways in which we ``image financial data``, i.e., transform a vector into a matrix. We provide specific details about each encoding method next.
\begin{enumerate}
 \item \textit{\textbf{Sequential Arrangement (SA).}} This is the first method we would like to examine. We simply take the sequential list of features within each accounting section and we create a long vector converted into a 2D image as illustrated in Figure \ref{Sequential_Arrangement}.
    \item \textit{Random Arrangement (RA).}\label{RA} To be able to provide a statistical analysis of the results we randomly rearrange the sequence of the features into a $486 \times 1$ vector, and then convert it into a $18 \times 27$ matrix which serves as the input image.
    \item \textit{\textbf{Category Chunk Arrangement (CCA).}} This is the second of our methods. For each financial statement section we add zeroes to obtain a $9 \times 9$ squared matrix. We concatenate all matrices as in Fig.~\ref{Categorical Chunk Arrangement}.
\item \textit{Within Chunk Randomization (WCR).}\label{WCR} On top of CCA, we randomize the features within each chunk. We keep the same chunk structure as in Fig.~\ref{Categorical Chunk Arrangement} 
\item \textit{Between Chunk Randomization (BCR).}\label{BCR} Based on CCA, we randomly shuffled the position of the six $9 \times 9$ chunks in the larger matrix. There are actually $6!$ possible arrangements. 
\item \textbf{\textit{Hilbert Vector Arrangement (HVA).}} In this encoding we follow the methodology described in \cite{hilbert1935stetige}, using Hilbert space-filling curves, illustrated in Fig. \ref{Hilbert Curve}. The authors argue and prove that using this encoding method preserves data clusters that may exist in the original data \cite{moon2001analysis}. 
\item \textit{Hilbert Vector Randomization (HVR).}\label{HVR} Since the Hilbert Vector Arrangement gives us squared images that are different from the rectangular image representations listed above, we design a control group by randomizing the Hilbert Vector Arrangement. We shuffle the sequenced features and then fill in the squared matrices following the Hilbert curve representation.
\end{enumerate}

\paragraph{Sequential Arrangement}
 Let  $\vec{x} = (x_{1}, x_{2}, \dotsm, x_{d})  \in  {\mathbb R}^{d}$ be a feature set with $d$ features. We simply take the original $d$- dimensional vector of features and transform it into a matrix $\vec{x} \in  {R}^{(a\times b)}$ by adding zeros at the end. Here we need $d \leq a\times b$ and in practical application we use $18\times 27$ to cover the $332$ fundamental data and we use $8\times 16$ to cover the $69$ financial ratios. Figure.\ref{Sequential_Arrangement} graphs this arrangement. 

\begin{figure}
\centering
\[
\begin{bmatrix} 
x_{1} & x_{2} & \dots & x_{b}\\
x_{b+1} & x_{b+2} & \dots & x_{2\times b}\\
\vdots & \vdots & \ddots & \vdots\\
x_{(a-1)\times b + 1} & x_{(a-1)\times b + 2} & \dots & x_{a \times b} 
\end{bmatrix}
\]
\caption{Sequential Arrangement for an $a \times b$ matrix\label{Sequential_Arrangement}}
\end{figure}

\begin{figure}[htbp]
\centerline{\includegraphics[width=8.8cm]{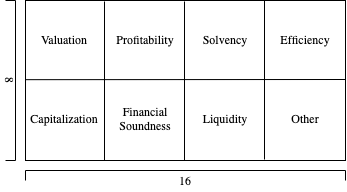}}
\caption{Categorical Chunk Arrangement For Ratio Data}
\label{Chunk_Arrangement}
\end{figure}

\paragraph{Categorical Chunk Arrangement}
A second encoding method transforms the feature vector into a matrix form by segmenting the matrix into 6 smaller matrices, or "chunks".  Each chunk contains features from the same section as described in Table \ref{tab_AccountingFeature} for fundamental data and Table \ref{RatioFeature} for financial ratios. Fig.~\ref{Categorical Chunk Arrangement} illustrates the distribution of sectors as described in \cite{federal2014fasab}. Each chunk is a $9 \times 9$ matrix containing the respective category padded with zeros in case there are insufficient variables to fill the whole chunk. The Categorical Chunk Arrangement used for the financial ratio data is presented in Fig. \ref{Chunk_Arrangement}.

The reason for using such encoding methods and adding dimensionality to the accounting features is to utilize CNN's proven success in learning hierarchical feature representation \citep{hatami2018classification}. Image representation of the accounting items not only rebuilds the hierarchical relationship within a set of financial statement variables, but also allows CNN to filter neighboring items and retrieve local and global representation of the financial information. 

The drawback of this manual pre-encoding process of financial data into a matrix format is that it is slow and requires financial expertise. A possible solution is to design an automated encoding framework for the financial statements that combines accounting variables and facilitates a wider application. One such method is the Hilbert Vector Arrangement described next. 

\paragraph{Hilbert Vector Arrangement} According to the authors of \cite{moon2001analysis} ``Although Peano discovered the first space-filling curve, it was Hilbert, in 1891, who was the first to recognize a general geometric procedure that allows the construction of an entire class of space-filling curves''. This method uses fractals to encode an 1-dimensional vector into a higher dimensional space. In figure \ref{Hilbert Curve} we see an example of encoding a 1 dimensional vector into a 2 dimensional image. 

The encoding is designed to preserve the distance between the data points which are close to each other and reduce the distance between points which are far away from each other. A Hilbert curve requires the vector to be transformed into a squared matrix in the 2-D space with the side length $d = 2^n$ for n-th order (e.g. $d=2$ for the first order Hilbert curve, $d=4$ for the second order Hilbert curve, etc.). With a side length $d$, the squared matrix can hold up to $d^2$ features. In our implementation we use a square matrix of $32 \times 32$ for the 332 fundamental features, and a square matrix of $16 \times 16$ for the 69 financial ratio features. After we fill the matrix with features, we complete the matrix with zeros.

\paragraph{Various Randomization of Schemes}
To obtain a credible and robust conclusion on performance, we have to compare the results obtain using our single encoding schemes with different types of randomized ``feature images''.
\highlight{Specifically, we use Data Encoding methods (\ref{RA}, \ref{WCR}, \ref{BCR}, \ref{HVR}) to randomly rearrange features into image, and we run 30 times for each randomization scheme to report their standard error.}
To this end, we create ''Random Arrangement'' and two different types of chunk randomization. We also create a Hilbert Vector Randomization. 
We should mention that the difference between the Hilbert Vector Randomization and the Random Arrangement is mainly the proportion of features versus zeros in the final matrices. This is caused by the required size differences of the matrices, since the Hilbert matrix has to be a square matrix.


\subsection{CNN}

\begin{figure}[htbp]
\centerline{\includegraphics[width=8.8cm]{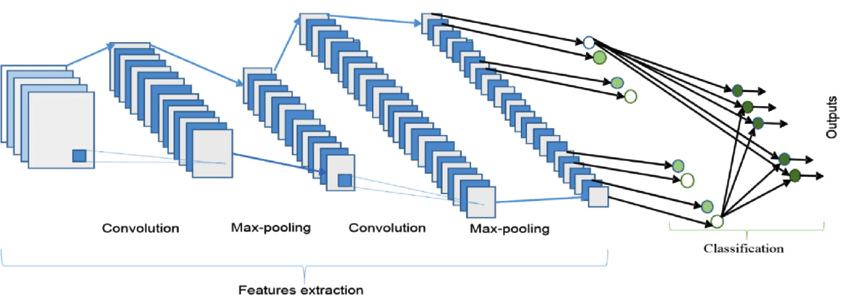}}
\caption{A General CNN Structure}
\label{cnn}
\end{figure}

It is widely accepted that Convolutional Neural Network (CNN) is a successful method for classification when applied to computer vision and text recognition problems \citep{driss2017comparison, lecun1995convolutional}. 
A general CNN structure includes convolutional layers for feature learning and selection, as well as dense layers for classification. An illustration of the architecture may be found in Figure \ref{cnn}. 

For image recognition and classification, CNN assumes that local connections exist between the input vector data points. Using filters with a specific size, CNN can capture such connections from a 2D image. With a stride of $k$, the filters move along the input tensor $k$ points at a time, and extract relevant features by applying convolutional operations to the input. A drawback for the convolutional filters is that existing information along the  edges of the images may be neglected. A fix for this is to pad 0's in several rows and columns around the edges for a better performance.
 
Pooling layers, such as max pooling and average pooling, is a sample-based discretization process. The purpose is to deal with high dimensional data and reduce the dimensionality of the input feature set. A combination of convolution and pooling layers can capture the temporal dynamics of time series, according to \cite{tsantekidis2017forecasting}. After the last set of convolution/pooling layers, a set of fully connected dense layers are used to perform the final classification task. 

In this paper, the output layer contains the distinct classes of the corporate credit rating with one-hot encoding. The error in prediction is obtained by applying a categorical cross entropy loss function on the output layer. CNN is trained using a classic back-propagation algorithm based on minimizing the categorical cross entropy loss function \citep{ werbos1990backpropagation}.






Comparing to a standard feed-forward neural network, CNN is considered an easier training method due to the lesser number of parameters and connections. CNN is also capable of selecting useful financial features during the training process as a result of convolution operations \citep{di2016artificial}.

In this work, we implement two different CNN models to fit the 1-dimensional and 2-dimensional input data.
\highlight{We use a grid search method to tune the number of neurons in each CNN layer (refer to Appendix \ref{app:grid_search} for the details).}
The 1D convolution neural network architecture contains two 1D convolution layers, with 64 and 32 units in each layer. The filter window is size 3. The 2D convolution neural network architecture has two convolution layers, with 64 and 32 units in each layer. The filter size is 3 by 3. Both convolutional architectures are followed by two fully connected layers with 128 nodes each.


\section{Empirical Results} \label{sec:results}
In this section we are analyzing the results obtained using the different ways of imaging data. 

\subsection{Benchmark model and Performance indicator measures}

To compare the performance of our encoding methods with non-feature learning methods, we choose a two-layer Multilayer Perceptron (MLP) with dropout rate 0.3 \citep{srivastava2014dropout} as a benchmark model. 

To get a credible and robust conclusion on the performance of our model, we select two indicators of model performance. We use a traditional \textbf{prediction accuracy} as well as a \textbf{`Notch Distance`}, introduced in \cite{parisa2020}. The accuracy is simply the proportion of the ratings correctly classified according to S\&P ratings. The Notch Distance detailed in equation \ref{notchdis} measures the distance between the S\&P rating and the model prediction. 
\begin{equation}\label{notchdis}
    \mathbf E[|Y|] = \sum_{i} |i| \times \text{F(i)}.
\end{equation}
Here, $Y$ is a random variable representing the difference between the S\&P rating and the model prediction, and $F$ is the \textit{frequency} of each difference (also called notch) $i$. The difference corresponds to each point in the test dataset $i \in \{\cdots,-1,0,1,\cdots\}$ and is dependent on the particular encoding used. The equation \ref{frequency} formalizes the frequency definition:
\begin{equation}\label{frequency}
  \text{F}(i) = \sum_{k \in N} I_{(\hat{y}_k - y_k = i)} /N,
\end{equation}
where $N$ is the total number of observations in the test set, $I$ is the indicator function. Here, $y$ denotes the true rating of an observation, and $\hat y$ denotes the prediction given by a particular model.

The authors of \cite{parisa2020} argue that the measure is influenced by the percentage of the correct predictions (value $0$). To have a more accurate measure in terms of average notch distance when the prediction is wrong, we can calculate a conditional expectation:
$$\mathbf E[Y\mid Y\neq 0] = \sum_{i\neq 0} i \times \frac{\text{F(i)}}{\sum_{j\neq 0 }\text{F(j)}}$$
This conditional expectation eliminates the correct predictions and its expectation should be a better measure of how many notches we expect the algorithm to be off when the prediction fails. Said conditional expectation is reported in all tables as the \textbf{Notch Distance}. 

\subsection{Performance Evaluation}\label{perf_res}

The major goal is to compare the different types of 2D encoding methods. To have an estimate for the randomness of the results, we create the randomized 2D images as described in the previous section. We randomize each input type 30 times to obtain meaningful confidence intervals. We denote the results obtained for these randomized arrangements with italic letters in the following tables. The numbers within parentheses in the table are standard errors of the results. For the Notch distance a lower value is better.

\begin{table}[htbp]
  \centering
  \caption{Comparison of Accuracy and Notch Distance with different encoding methods for Financial sector (Test Period: 2016)}
    \begin{tabular}{lll}
    \hline
          & Accuracy & Notch Distance \\
          \hline
    MLP   & 0.452 & 1.576 \\
    1D CNN & 0.535 & 1.545 \\
    \hline
    Sequential Arrangement (SA) & 0.390* & 1.544 \\
    \textit{Random Arrangement} & {0.348 (0.007)} & {1.668 (0.02)} \\
    Category Chunk Arrangement (CCA) & 0.444* & 1.724 \\
    \textit{Within  Chunk  Randomization} & {0.397 (0.006)} & {1.673 (0.021) }\\
   \textit{{Between Chunk Randomization}} & 0.404 (0.006) & 1.725 (0.025) \\
    Hilbert Vector Arrangement (HVA) & 0.461* & 1.523 \\
    \textit{Hilbert Vector Randomization} & 0.363 (0.009) & 1.592 (0.017) \\
    \hline
    \end{tabular}%
    \label{tab:fin_perf}%
    \begin{tablenotes}
      \small
      \item *significant at $p < 0.05$
    \end{tablenotes}
  
\end{table}%

\begin{table}[htbp]
  \centering
  \caption{Comparison of Accuracy and Notch Distance with different encoding methods for Healthcare sector (Test Period: 2016)}
    \begin{tabular}{lll}
    \hline
          & Accuracy & Notch Distance \\
          \hline
    MLP   & 0.620 & 1.431 \\
    1D CNN & 0.608 & 1.373 \\
    \hline
    Sequential Arrangement  (SA) & 0.620* & 1.677 \\
    \textit{Random Arrangement} & 0.476 (0.0093) & 1.724 (0.037) \\
    Category Chunk Arrangement (CCA) & 0.602* & 1.500 \\
    \textit{Within  Chunk  Randomization} & 0.575 (0.0081) & 1.539 (0.0259) \\
    \textit{Between Chunk Randomization} & 0.557 (0.0059) & 1.514 (0.0229) \\
    Hilbert Vector Arrangement (HVA) & 0.585* & 1.437 \\
    \textit{Hilbert Vector Randomization} & 0.471 (0.0056) & 1.732 (0.0343) \\
    \hline
    \end{tabular}%
  \label{tab:hea_perf}%
  \begin{tablenotes}
      \small
      \item *significant at $p < 0.05$
    \end{tablenotes}
\end{table}%

\begin{table}[htbp]
  \centering
  \caption{Comparison of Accuracy and Notch Distance with different encoding methods for IT sector (Test Period: 2016)}
    \begin{tabular}{lll}
    \hline
          & Accuracy & Notch Distance \\
          \hline
    MLP   & 0.490 & 2.000 \\
    1D CNN & 0.574 & 1.747 \\
    \hline
    Sequential Arrangement (SA)  & 0.485 & 1.867 \\
    \textit{Random Arrangement} & 0.488 (0.0054) & 1.775 (0.0248) \\
    Category Chunk Arrangement (CCA)& 0.515 & 1.919 \\
    \textit{Within  Chunk  Randomization} & 0.533 (0.0062) & 1.746 (0.0236) \\
    \textit{Between Chunk Randomization} & 0.508 (0.0049) & 1.803 (0.0216) \\
   Hilbert Vector Arrangement (HVA) & 0.485 & 1.838 \\
    \textit{Hilbert Vector Randomization} & 0.514 (0.0043) & 1.765 (0.0113) \\
    \hline
    \end{tabular}%
  \label{tab:it_perf}%
  \begin{tablenotes}
      \small
      \item *significant at $p < 0.05$
    \end{tablenotes}
\end{table}%

\textbf{About evaluation methods.} The two methods of evaluating are based on accuracy and conditional notch distance. The two measures are outputting different insights about a method performance. The accuracy is very clear - it reflects the percent of times that the respective model outputs the corrects rating in the test data. Correct here is defined by the S\&P ratings. The notch distance tells us how far the prediction was on average from the correct result, given that the output of the model was incorrect. So for example, in the financial sector (Table \ref{tab:fin_perf}) using CCA, when the prediction was wrong, it was about 1.7 notches away from the correct rating (according to our encoding method). This allows us to compare the accuracy performance, as well as assess how far the model was from the correct result. We see the notch distance as a second level of comparison. Referring to the same Table \ref{tab:fin_perf} for example, the HVA is more accurate than SA. However, CCA and HVA have similar accuracy numbers, and thus to decide between them we look at the notch distance to see that whenever HVA was wrong it was ``less wrong'' than CCA. 

\highlight{We did not use recall or the f1 score as an evaluation measure. F1 score is good when both precision and recall are important. However, this is not our case. For example, to monetize our study, a portfolio manager would invest in those companies whose ratings are predicted to upgrade compared with the previous quarter. In such case, recall and f1 score are not as important as precision.  To exemplify, suppose our model predicts 10 companies being upgraded soon, and 9 out of 10 are correctly predicted. Thus, the precision is 0.9. At the same time, there are 100 companies actually being upgraded within the next quarter, thus the recall and f1 scores are 0.09 and 0.16, respectively. The model will be evaluated as a bad model based on the low f1 score, but actually it is a great and profitable model because we invest 90\% of our wealth in those companies whose stock price most likely would go up. Considering another model, where 5 out of 10 are correctly predicted, and there are 6 companies actually upgraded in the market within the next quarter. The precision, recall and f1 score are 0.5, 0.83 and 0.62. The f1 score indicates a better model but actually it is a less profitable model than the first one. For completeness, we added recall and f1 scores in Appendix \ref{pre_f1}.}

Tables \ref{tab:fin_perf}, \ref{tab:hea_perf} and \ref{tab:it_perf}
present the\highlight{mean and standard error for} accuracy and notch distance for the financial, healthcare and IT sector respectively. These results use the fundamental data directly. 
Tables \ref{tab:ratio_fin}, \ref{tab:ratio_hea}, \ref{tab:ratio_it} present the\highlight{mean and standard error for} accuracy and notch distance for the same sectors however we use the the financial ratios as input variables (instead of the raw fundamental variables). 
In theory financial ratios are more meaningful for comparison between company statements. They are supposed to reveal insights regarding profitability, liquidity, operational efficiency, and solvency.

\begin{table}[htbp]
  \centering
  \caption{Comparison of Accuracy and Notch Distance for financial ratios with different encoding methods for Financial sector (Test Period: 2016)}
    \begin{tabular}{lll}
    \hline
          & Accuracy & Notch Distance \\
          \hline
    MLP   & 0.468 & 1.575 \\
    1D CNN & 0.503 & 1.566 \\
    \hline
    Sequential Arrangement  (SA) & 0.555* & 1.678 \\
    \textit{Random Arrangement} & 0.537 (0.004) & 1.438 (0.017) \\
    Category Chunk Arrangement (CCA)& 0.549* & 1.630 \\
    \textit{Within  Chunk  Randomization} & 0.513 (0.006) & 1.627 (0.022) \\
    \textit{Between Chunk Randomization} & 0.507 (0.005) & 1.464 (0.018) \\
    Hilbert Vector Arrangement (HVA) & 0.536* & 1.376 \\
    \textit{Hilbert Vector Randomization} & 0.516 (0.004) & 1.482 (0.023) \\
    \hline
    \end{tabular}%
  \label{tab:ratio_fin}%
  \begin{tablenotes}
      \small
      \item *significant at $p < 0.05$
    \end{tablenotes}
\end{table}%

\begin{table}[htbp]
  \centering
  \caption{Comparison of Accuracy and Notch Distance for financial ratios with different encoding methods for Healthcare sector (Test Period: 2016)}
    \begin{tabular}{lll}
    \hline
          & Accuracy & Notch Distance \\
          \hline
    MLP   & 0.678 & 1.883 \\
    1D CNN & 0.634 & 2.126 \\
    \hline
    Sequential Arrangement  (SA)  & 0.695 & 2.315 \\
    \textit{Random Arrangement} & 0.688 (0.0048) & 1.862 (0.0235) \\
    Category Chunk Arrangement (CCA)& 0.663 & 1.745 \\
    \textit{Within  Chunk  Randomization} & 0.664 (0.0048) & 1.912 (0.0282) \\
    \textit{Between Chunk Randomization} & 0.675 (0.0043) & 1.992 (0.0294) \\
    Hilbert Vector Arrangement (HVA) & 0.655 & 1.976 \\
    \textit{Hilbert Vector Randomization} & 0.685 (0.0069) & 1.923 (0.0421) \\
    \hline
    \end{tabular}%
  \label{tab:ratio_hea}%
  \begin{tablenotes}
      \small
      \item *significant at $p < 0.05$
    \end{tablenotes}
\end{table}%

\begin{table}[htbp]
  \centering
  \caption{Comparison of Accuracy and Notch Distance for financial ratios with different encoding methods for IT sector (Test Period: 2016)}
    \begin{tabular}{lll}
    \hline
          & Accuracy & Notch Distance \\
          \hline
    MLP   & 0.564 & 2.046 \\
    1D CNN & 0.673 & 2.071 \\
    \hline
    Sequential Arrangement  (SA)  & 0.686* & 2.158 \\
    \textit{Random Arrangement} & 0.661 (0.0043) & 2.015 (0.039) \\
    Category Chunk Arrangement (CCA) & 0.665 & 2.113 \\
    \textit{Within  Chunk  Randomization} & 0.691 (0.0032) & 2.090 (0.0299) \\
    \textit{Between Chunk Randomization} & 0.676 (0.0045) & 2.060 (0.0242) \\
    Hilbert Vector Arrangement (HVA) & 0.660 & 2.272 \\
    \textit{Hilbert Vector Randomization} & 0.675 (0.0038) & 2.130 (0.0343) \\
    \hline
    \end{tabular}%
  \label{tab:ratio_it}%
  \begin{tablenotes}
      \small
      \item *significant at $p < 0.05$
    \end{tablenotes}
\end{table}%

Reading these tables we are particularly interested in answering three questions. 
\begin{enumerate}
    \item How do the numbers obtained for a particular arrangement compare with the numbers obtained if the arrangement is randomized? To ease this comparison we put the randomized numbers for each method directly under each method. We use a simple t-test for this comparison.
    \item Is there a consistently best performer among the different 2D imaging techniques? We perform pairwise t-tests with a Bonferroni correction for this comparison.
    \item How does the best performer among the 2D imaging techniques compares with results obtained for just 1 dimensional methods (MLP and 1D CNN)? 
\end{enumerate}

\highlight{We shall answer these questions one by one in the next few paragraphs.
We use the acronyms SA, CCA, HVA, MLP, and 1d CNN or simply CNN to refer to the different imaging methods used.}

\textbf{Comparing performance for the 2D encoding methods and their randomization.} 
\highlight{We perform one sided t-tests for each method and compare with their randomized version. The formal hypotheses are:}
\highlight{
\begin{small}
\begin{align*}
& H_0: \text{The accuracy for a specific encoding (SA, CCA, HVA) is} \\
& \text{equal to the mean accuracy for the respective randomization scheme} \\
& H_a: \text{The results for the 2D encoding method are better than those} \\
& \text{obtained for the respective randomization scheme}
\end{align*}
\end{small}
}
The variability in the testing procedure here comes entirely through rearranging the variables. The data split (training/test) is the same for each result. 

The results are inconclusive for this question. For example, HVA, SA, and CCA are significantly better than their random arrangements for the financial sector. This is reflected in both fundamental data and financial ratios data. However, in the IT sector they are not different from random arrangements. Looking at the healthcare sector, HVC, SA, and CCA show significantly different accuracy numbers from their random arrangement when using fundamental data as input (Table \ref{tab:hea_perf})  but show no significant difference when using financial ratios as input (Table \ref{tab:ratio_hea}). The conclusion may be dependent on sector. For the financial sector the particular categorization of accounting variables we describe in Section \ref{sec:dataRestructuring} seems to be important, while for IT seems not relevant. 

\textbf{Comparing the performance of the three 2D encoding methods.}
There seems to be no method that outperforms other 2D encoding techniques. As mentioned before we compare the results using a pairwise t-test with Bonferroni correction \citep{bland1995multiple}, and we summarized the results of all the tests in Table \ref{tab:rank_1}. 

\highlight{
Formally the test has hypotheses: 
\begin{small}
\begin{align*}
& H_0: \text{The accuracy for all encoding methods (SA, CCA, HVA)} \\
& \text{is the same} \\
& H_a: \text{The accuracy for the 2D encoding methods is different} 
\end{align*}
\end{small}
}

The values within the same circle are not statistically different. Looking at the Table \ref{tab:rank_1}, we note that contrary to our expectations HVA generally is among the worst performing imaging technique, however there is no clear winner. A few things in common for all the sectors are that: CCA is always in the top performing group when using Fundamental Data as input, while SA performs better when using financial ratio data. Thus if we were to recommend a technique, the choice would be CCA when using fundamental data as input, and SA when using financial ratio data. 
\begin{table}[htbp]
  \centering
  \caption{pairwise t-tests with Bonferroni correction. Lower rank is better. Circled values are not significantly different}
     \begin{tabular}{cccc}
     \hline
    Ranking & 1 & 2 & 3\\
    \hline\hline
    Financial Sector + Fundamental Data & \tikzmark{startup}HVA & CCA\tikzmark{endup} & SA \\
    \hline
    Financial Sector + Ratio Data & \tikzmark{startup1}SA & \tikzmark{startup2} CCA\tikzmark{endup1} & HVA\tikzmark{endup2} \\
    \hline\thickhline
    Healthcare Sector + Fundamental Data & \tikzmark{startup3}SA & \tikzmark{startup4} CCA\tikzmark{endup3} & HVA\tikzmark{endup4} \\
    \hline
    Healthcare Sector + Ratio Data & SA &  \tikzmark{startup5}CCA & HVA\tikzmark{endup5}\\
    \hline\thickhline
    IT Sector + Fundamental Data & CCA & \tikzmark{startup6}HVA & SA\tikzmark{endup6} \\
    \hline
    IT Sector + Ratio Data & SA & \tikzmark{startup7}CCA & HVA \tikzmark{endup7}\\
    \hline
    \end{tabular}%
  \label{tab:rank_1}%
  
  \begin{tikzpicture}[remember picture,overlay]
\foreach \Val in {up, up1, up2, up3, up4,up5, up6, up7}
{
\draw[rounded corners,black,thick]
  ([shift={(-0.5\tabcolsep,-0.5ex)}]pic cs:start\Val) 
    rectangle 
  ([shift={(0.5\tabcolsep,2ex)}]pic cs:end\Val);
}
\end{tikzpicture}

\end{table}%

\textbf{Comparison of Performance for 2D encoding methods versus 1D methods.}
There is no clear winner for this comparison either. To elaborate, when using Fundamental Data, the best 1D method is statistically more accurate  than the best performer among the 2D methods (Tables \ref{tab:fin_perf}, \ref{tab:hea_perf}, \ref{tab:it_perf}).
However, the results are reversed when we use Financial Ratio Data as input (Tables \ref{tab:ratio_fin}, \ref{tab:ratio_hea}, \ref{tab:ratio_it}). From these results it would seem that when working with Fundamental data a 2D re-imaging is unnecessary. When using financial ratio data it would appear that reformatting the data into a 2D image would produce better results. We will investigate possible reasons for this dichotomy in the next section. 

Comparing the results for fundamental data (Tables \ref{tab:fin_perf}, \ref{tab:hea_perf}, \ref{tab:it_perf}) with the financial ratio data results (Tables \ref{tab:ratio_fin}, \ref{tab:ratio_hea}, and \ref{tab:ratio_it}), we see an improvement in accuracy. However, the notch distance increases as well. Recall that the notch distance measures how far the prediction is from the true value given that the prediction is wrong.  Although Financial Ratios improve model accuracy, if we look at the notch distance values we see that: when prediction is wrong it is further away from the true value then when using fundamental data. Thus, using Financial ratio data may potentially have drawbacks for investors who make decisions based on corporate credit rating assessment.

\textbf{In conclusion, what is the best way to encode the input data?}
Combining the answers of the previous questions, we believe it is better to apply a 1D method on Fundamental Data without any imaging. If one uses  Financial Ratios to assess credit worthiness one should use a simple SA method for imaging data.

In the next section we investigate possible causes why 2D imaging did not have the significant impact we expected. In particular, why 2D imaging works for financial ratios and does not work for fundamental data. 
 
\subsection{Investigating causes why imaging is not universally better}

When transforming the original data into a matrix, we had to add $0$ values. The reason is that we had to transform data into a square or rectangle to be treated as an image and the original number of variables did not fit this pattern. These $0$ values may have introduced noise into the model. For example, when creating the Category Chunk Arrangement we added $692$ zeroes for fundamental data. However, when using financial ratios for CCA we only added 59 zeroes. It may be the truth that the lesser the number of zeros contributed to imaging methods, the better the performance for financial ratios.

Another possible cause of the poor performance when using fundamental data, could be the fact that financial ratios are already a combination of original accounting variables. The way the data is transformed from fundamental to financial ratio data may be similar to the way auto-encoders are supposed to work in artificial neural networks \citep{barnes1987analysis}.  Specifically, auto-encoders are supposed to learn a representation (encoding) of a data set, typically for dimensionality reduction, by training the network to ignore signal “noise”. 

In this section, we are further asking two more questions:
\begin{enumerate}
\item Does padding with zeros have any influence on the performance of 2D encoding?
\item Would using an `auto-encoder' on fundamental data instead of financial ratios improve the performance?
\end{enumerate}

\textbf{Effect of padding with zeroes} To answer this question, we implement a \textit{Reduced-Zero Padding} encoding method. Specifically, we use a reduced numbers of features so that the resulting number of features may be put into a squared matrix without adding any zeros. The features dropped are the ones that contained most missing values. The fundamental data feature set is reduced to 256 (from 332) for fundamental data and to 64 (from 69) for financial ratios data. We then use the Hilbert vector method to encode data as input for the 2D CNN model. We use HVA as benchmark because transforming data into a squared matrix is much easier than the rectangular shapes needed for CCA and SA. 

Table \ref{tab:reduce} presents the results obtained when using HVA. When we have no zeroes padded the results are improved in certain cases and made worse in several cases. This indicates that the padding with zeros is not a significant factor impacting the performance of 2D encoding.

\begin{table}[htbp]
  \centering
  \caption{Performance of Non-zero padding encoding}
    \begin{tabular}{|c|c|cc|}
    \hline
    \multicolumn{1}{|c}{} & \multicolumn{1}{c|}{} & HVA Accuracy using & HVA Original\\
    \multicolumn{1}{|c}{} & \multicolumn{1}{c|}{} & Reduced-zero Padding& Accuracy \\
    \hline
     & Fundamental data & 0.398 & 0.461 \\
     {Financial}     & Financial Ratios & 0.546 & 0.536 \\
          \hline
     & Fundamental data & 0.608 & 0.585 \\
    {Healthcare}      & Financial Ratios & 0.701 & 0.655 \\
          \hline
     & Fundamental data & 0.529 & 0.485 \\
     {IT}     & Financial Ratios & 0.654 & 0.660 \\
          \hline
    \end{tabular}%
  \label{tab:reduce}%
\end{table}%

\textbf{Effect of using an `auto-encoder'} 
To perform this experiment we implement the auto-encoder architecture from \cite[Table 2]{bahadur2019dimension}. This helps us encode the original 332 fundamental features into 69 output variables. We use 69 as that is the number of financial ratios and we wanted to have a direct comparison. 

Table \ref{tab:ae} compares the results obtained from the ``auto-encoder'' with results obtained using the SA model. The accuracy of the auto-encoder model is worse even when compared with the performance obtained using the original fundamental variables. 

This implies that financial ratios are indeed well thought numbers and today's computer ``auto-summarizing'' features do not come close to them.

\begin{table}[htbp]
  \centering
  \caption{Auto-encoder method results compared with the results of SA on fundamental data}
    \begin{tabular}{c|ccc}
    \hline
    \multicolumn{1}{c}{} & Auto-encoder  &  Accuracy  & Accuracy  \\
    \multicolumn{1}{c}{} & Accuracy &  (Fundamental data) &  (Financial ratios) \\
    

    \hline
     {Financial}     
           & 0.183 & 0.390 & 0.555\\
          \hline
    {Healthcare}      
          & 0.064 & 0.620 & 0.695\\
          \hline
     {IT}     
           & 0.113 & 0.485 & 0.686\\
          \hline
    \end{tabular}%
  \label{tab:ae}%
\end{table}%

\section{Conclusion}

In this work we investigate whether encoding financial data into an image format is beneficial for assessing corporate credit rating. We implement three encoding methods (CCA, SA, HVA) and we use them to answer several questions relevant to image encoding. The results obtained provide several recommendations about formatting the input data to be used in a convolutional network architecture.

Several recent papers point to the Hilbert Vector Arrangement as the best method to encode 1D data into a 2D image input \citep{kavitha2017deep, yin2018image, loka2019hilbert}. Our experiments with financial data did not confirm this assessments and we found no universally best encoding technique. Specifically, we found Category Chunk Arrangement as an adequate way to encode 1D data into a 2D image when using fundamental data, and Sequential Arrangement as an appropriate technique when using financial ratio data.

More importantly, we found that these imaging techniques were sometimes inferior to machine learning techniques applied directly to 1D data. This is particularly true when using fundamental data as input. Thus, we recommend using a regular 1D CNN when analyzing fundamental data. When using financial ratio data as input, the imaging methods are worth performing and produce significantly better results. Our recommendation is to use a Sequential Arrangement as it is straightforward and produced top results every time we used it. 

We were puzzled by these results and we created two hypotheses. The dichotomy may be due to the zeroes we had to add to the data to create an input image. That is demonstrably not the case as we obtained similar results when we did not add zeroes to the data.  

The second item of interest was the reason why financial ratio data was producing so much better results. We hypothesised that an auto-encoder of fundamental data may produce similar performance with those obtained when using financial ratio data. We found out that the auto-encoder method we implemented produces much worse results than using the encoding provided by financial ratios. Thus, the hundreds of years of human experience encoding fundamental data into financial ratios is still superior to the way computer does it today.

\appendices
\section{Grid Search Results}\label{app:grid_search}

\begin{table}[htbp]
  \centering
  \caption{\highlight{Grid Search Performance}}
    \begin{tabular}{cccc}
    \hline
    Training Accuracy & Test Accuracy & neurons1 & neurons2 \\
    \hline
    0.920 & 0.527 & 16    & 16 \\
    0.896 & 0.523 & 16    & 32 \\
    0.908 & 0.523 & 16    & 64 \\
    0.881 & 0.564 & 16    & 128 \\
    0.900 & 0.490 & 32    & 16 \\
    0.905 & 0.552 & 32    & 32 \\
    0.912 & 0.539 & 32    & 64 \\
    0.920 & 0.481 & 32    & 128 \\
    0.909 & 0.465 & 64    & 16 \\
    0.939 &\high{0.585}   &\high{64}    &\high{32} \\
    0.904 & 0.502 & 64    & 64 \\
    0.925 & 0.568 & 64    & 128 \\
    0.898 & 0.539 & 128   & 16 \\
    0.930 & 0.498 & 128   & 32 \\
    0.932 & 0.564 & 128   & 64 \\
    0.911 & 0.519 & 128   & 128 \\
    \hline
    \end{tabular}%
  \label{tab:grid_search}%
\end{table}%

\section{Precision and f1 score}\label{pre_f1}

\begin{table}[htbp]
  \centering
  \caption{\highlight{Comparison of Precision and F1 score with different encoding methods for Financial sector (Test Period: 2016)}}
    \begin{tabular}{lll}
    \hline
          & Precision & f1 score  \\
    \hline
    Sequential CNN & 0.675 & \highli{0.520} \\
    \hline
    Sequential Arrangement & 0.518 & 0.345 \\
    Random Arrangement & 0.389 (0.012) & 0.288 (0.007) \\
    Category Chunk Arrangement & 0.511 & 0.390 \\
    Within  Chunk  Randomization & 0.441 (0.008) & 0.333 (0.007) \\
    Between Chunk Randomization & 0.448 (0.008) & 0.350 (0.007) \\
    Hilbert Vector Arrangement & 0.666 & 0.453 \\
    Hilbert Vector Randomization & 0.387 (0.013) & 0.299 (0.012) \\
    \hline
    \end{tabular}%
  \label{tab:fin_f1}%
\end{table}%

\begin{table}[htbp]
  \centering
  \caption{\highlight{Comparison of Precision and F1 score with different encoding methods for Healthcare sector (Test Period: 2016)}}
    \begin{tabular}{lll}
    \hline
          & Precision & f1 score  \\
    \hline
    Sequential CNN & 0.565 & 0.564 \\
    \hline
    Sequential Arrangement & 0.621 & \highli{0.587} \\
    Random Arrangement & 0.507 (0.008) & 0.455 (0.009) \\
    Category Chunk Arrangement & 0.606 & 0.581 \\
    Within  Chunk  Randomization & 0.559 (0.012) & 0.529 (0.009) \\
    Between Chunk Randomization & 0.556 (0.009) & 0.519 (0.007) \\
    Hilbert Vector Arrangement & 0.595 & 0.571 \\
    Hilbert Vector Randomization & 0.475 (0.008) & 0.450 (0.008) \\
    \hline
    \end{tabular}%
  \label{tab:hea_f1}%
\end{table}%

\begin{table}[htbp]
  \centering
  \caption{\highlight{Comparison of Precision and F1 score with different encoding methods for IT sector (Test Period: 2016)}}
    \begin{tabular}{lll}
    \hline
          & Precision & f1 score  \\
    \hline
    Sequential CNN & 0.51  & 0.446 \\
    \hline
    Sequential Arrangement & 0.441 & 0.376 \\
    Random Arrangement & 0.409 (0.008) & 0.380 (0.004) \\
    Category Chunk Arrangement & 0.559 & \highli{0.478} \\
    Within  Chunk  Randomization & 0.491 (0.007) & 0.433 (0.007) \\
    Between Chunk Randomization & 0.462 (0.008) & 0.413 (0.004) \\
    Hilbert Vector Arrangement & 0.468 & 0.403 \\
    Hilbert Vector Randomization & 0.426 (0.007) & 0.405 (0.003) \\
    \hline
    \end{tabular}%
  \label{tab:it_fi}%
\end{table}%

\begin{table}[htbp]
  \centering
  \caption{\highlight{Comparison of Precision and F1 score for financial ratios with different encoding methods for Financial sector (Test Period: 2016)}}
    \begin{tabular}{lll}
    \hline
          & Precision & f1 score \\
    \hline
    Sequential CNN & 0.615 & 0.528 \\
    \hline
    Sequential Arrangement & 0.591 & 0.536 \\
    Random Arrangement & 0.601 (0.009) & 0.550 (0.008) \\
    Category Chunk Arrangement & 0.573 & 0.525 \\
    Within  Chunk  Randomization & 0.580 (0.010) & 0.522 (0.009) \\
    Between Chunk Randomization & 0.599 (0.010) & 0.530 (0.008) \\
    Hilbert Vector Arrangement & 0.623 & \highli{0.579} \\
    Hilbert Vector Randomization & 0.579 (0.009) & 0.528 (0.008) \\
    \hline
    \end{tabular}%
  \label{tab:fin_ratio_f1}%
\end{table}%

\begin{table}[htbp]
  \centering
  \caption{\highlight{Comparison of Precision and F1 score for financial ratios with different encoding methods for Healthcare sector (Test Period: 2016)}}
    \begin{tabular}{lll}
    \hline
          & Precision & f1 score \\
    \hline
    Sequential CNN & 0.639 & 0.618 \\
    \hline
    Sequential Arrangement & 0.680  & \highli{0.656} \\
    Random Arrangement & 0.668 (0.005) & 0.653 (0.004) \\
    Category Chunk Arrangement & 0.635 & 0.636 \\
    Within  Chunk  Randomization & 0.650 (0.005) & 0.637 (0.004) \\
    Between Chunk Randomization & 0.679 (0.004) & 0.650 (0.004) \\
    Hilbert Vector Arrangement & 0.646 & 0.622 \\
    Hilbert Vector Randomization & 0.667 (0.007) & 0.650 (0.006) \\
    \hline
    \end{tabular}%
  \label{tab:hea_ratio_f1}%
\end{table}%

\begin{table}[htbp]
  \centering
  \caption{\highlight{Comparison of Precision and F1 score for financial ratios with different encoding methods for IT sector (Test Period: 2016)}}
    \begin{tabular}{lll}
    \hline
          & Precision & f1 score \\
    \hline
    Sequential CNN & 0.618 & 0.563 \\
    \hline
    Sequential Arrangement & 0.642 & \highli{0.613} \\
    Random Arrangement & 0.606 (0.004) & 0.546 (0.005) \\
    Category Chunk Arrangement & 0.614 & 0.588 \\
    Within  Chunk  Randomization & 0.635 (0.003) & 0.593 (0.005) \\
    Between Chunk Randomization & 0.601 (0.008) & 0.561 (0.008) \\
    Hilbert Vector Arrangement & 0.577 & 0.536 \\
    Hilbert Vector Randomization & 0.618 (0.004) & 0.563 (0.005) \\
    \hline
    \end{tabular}%
  \label{tab:it_ratio_}%
\end{table}%

\begin{table}[htbp]
  \centering
  \caption{pairwise t-tests with Bonferroni correction on precision. Lower rank is better. Circled values are not significantly different}
     \begin{tabular}{cccc}
     \hline
    Ranking & 1 & 2 & 3\\
    \hline\hline
    Financial Sector + Fundamental Data & HVA & \tikzmark{startup8}CCA & SA\tikzmark{endup8} \\
    \hline
    Financial Sector + Ratio Data & \tikzmark{startup9}HVA & \tikzmark{startup10} SA\tikzmark{endup9} & CCA\tikzmark{endup10} \\
    \hline\thickhline
    Healthcare Sector + Fundamental Data & \tikzmark{startup11}SA &  CCA & HVA\tikzmark{endup11} \\
    \hline
    Healthcare Sector + Ratio Data & SA &  \tikzmark{startup12}CCA & HVA\tikzmark{endup12}\\
    \hline\thickhline
    IT Sector + Fundamental Data & CCA & \tikzmark{startup13}HVA & SA\tikzmark{endup13} \\
    \hline
    IT Sector + Ratio Data & SA & CCA & HVA \\
    \hline
    \end{tabular}%
  \label{tab:rank_2}%
  
  \begin{tikzpicture}[remember picture,overlay]
\foreach \Val in {up8, up9, up10, up11, up12,up13}
{
\draw[rounded corners,black,thick]
  ([shift={(-0.5\tabcolsep,-0.5ex)}]pic cs:start\Val) 
    rectangle 
  ([shift={(0.5\tabcolsep,2ex)}]pic cs:end\Val);
}
\end{tikzpicture}

\end{table}%


\section*{Acknowledgment}

The authors would like to acknowledge the UBS research grant award to the Hanlon laboratories which provided partial support for this research.

\IEEEtriggeratref{8}
\IEEEtriggercmd{\enlargethispage{-5in}}

\ifCLASSOPTIONcaptionsoff
  \newpage
\fi


\bibliographystyle{chicago}
\bibliography{reference}

\begin{IEEEbiography}[{\includegraphics[width=1in,height=1.25in,clip,keepaspectratio]{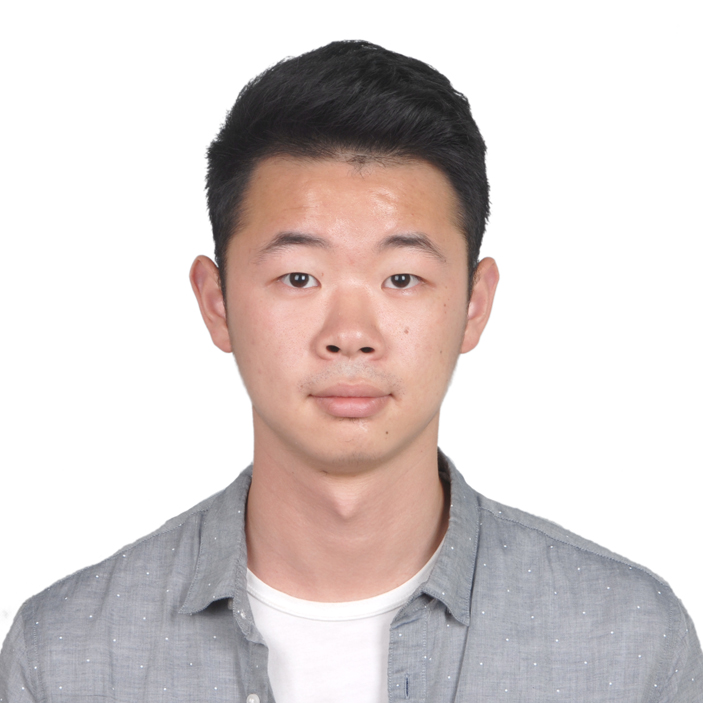}}]{Dan Wang}
was born in Zunyi, Guizhou, China in 1993. He received the B.S. degree in electronic engineering from Beihang university, Beijing in 2011 and the M.S. degree in financial engineering from Stevens Institute of Technology. He is currently pursuing the Ph.D. degree in Financial Engineering at Stevens Institute of Technology, Hoboken, NJ, USA.

From 2018 to present, he was a Research Assistant with Hanlon Financial Systems Lab. His research interest includes machine learning and deep learning application in finance, high frequency corporate credit rating, feature selection for financial statement, and volatility spillover in global markets.

\end{IEEEbiography}

\begin{IEEEbiography}[{\includegraphics[width=1in,height=1.25in,clip,keepaspectratio]{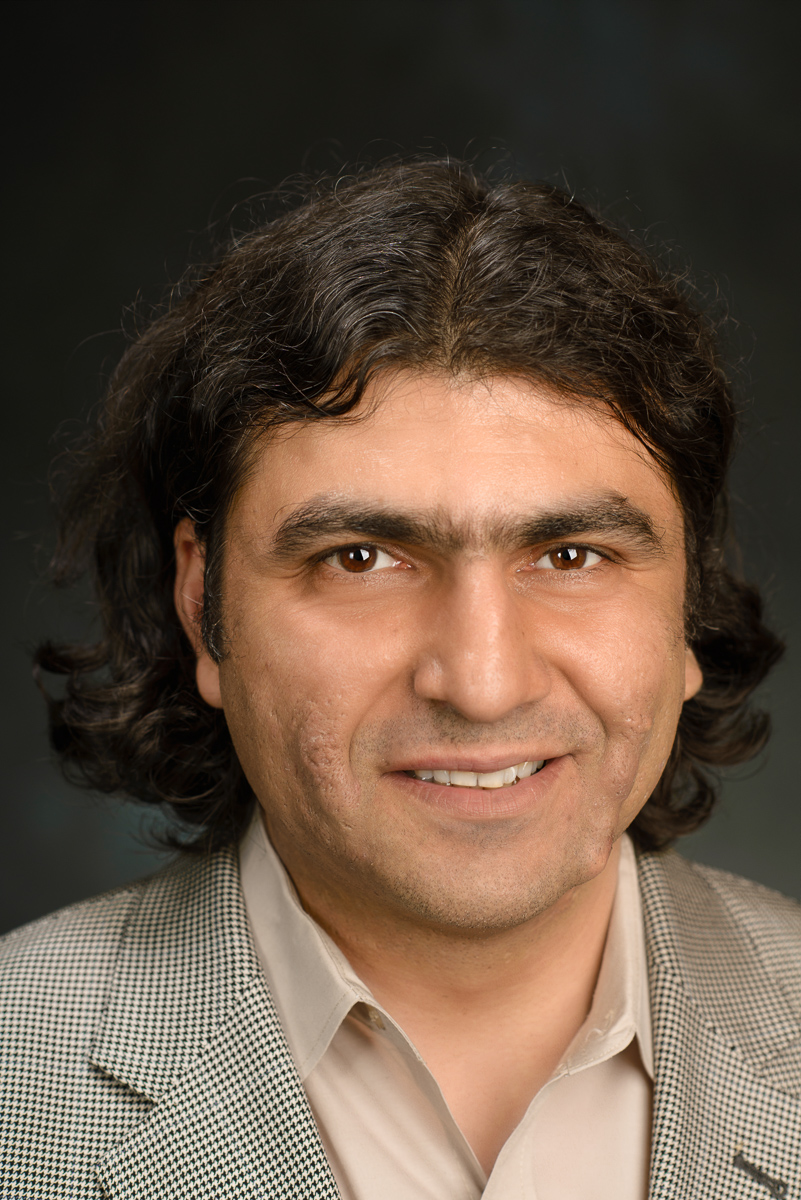}}]{Ionut Florescu}
Dr. Florescu is a research professor and director of both the Financial Analytics program and the Hanlon Financial Systems Laboratories. He has written four books and served as an editor for four more, has published more than 50 peer-reviewed publications, and has given presentations at conferences and seminars worldwide. Dr. Florescu is the principal investigator for five NSF awards and many other grants, and has a U.S. patent in computer vision.
\end{IEEEbiography}

\begin{IEEEbiography}[{\includegraphics[width=1in,height=1.25in,clip,keepaspectratio]{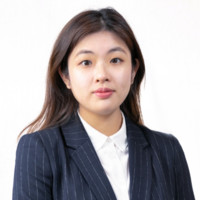}}]{Tianrui Wang}
was from Taiyuan, Shanxi, China. She received her Bachelor's degree in Accountancy from Hong Kong Polytechnic University in 2018, and is pursuing her Master's degree in Business Intelligence \& Analytics with a concentration in Data Science from Stevens Institute of Technology, Hoboken, NJ, USA. She has a keen interest in the application of machine learning and deep learning methods in the finance industry. Since May 2019, she has been conducting research on Credit Rating Prediction under the supervision of Dr. Florescu. 
\end{IEEEbiography}


\vfill


\end{document}